\documentstyle[sprocl,epsbox]{article}

\def\Journal#1#2#3#4{{#1} {\bf #2}, #3 (#4)}

\def\NPB{{\em Nucl. Phys.} B}
\def\NPBP{{{\em Nucl. Phys.}{(Proc. Suppl.)}} B}

\def\be{\begin{equation}}
\def\ee{\end{equation}}
\def\bea{\begin{eqnarray}}
\def\eea{\end{eqnarray}}

\begin{document}

\title{BEHAVIOR~OF~MONOPOLE~TRAJECTORIES
 IN THE~INSTANTON AND~ANTI-INSTANTON~SYSTEM AT~FINITE~TEMPERATURE}

\author{F. Araki, H. Suganuma and H. Toki}

\address{Research Center for Nuclear Physics, Osaka University\\
Mihogaoka 10-1, Ibaraki, Osaka 567, Japan}


\maketitle\abstracts{ 
We study the connection between monopoles and instantons 
in the Polyakov-like gauge,
and the behavior of the monopole trajectories generated by instantons 
at finite temperature.
The monopole trajectories become long and complicated
at the zero-temperature which can be regarded as monopole condensation. 
On the other hand, they tend to become simple
at finite temperature.
It is suggested that such a change is related to 
the confinement-deconfinement phase transition.}

\section{Introduction}

The nonabelian gauge theory like QCD is reduced to the abelian gauge theory 
including monopoles under the abelian gauge fixing \cite{thooft}.
It is thought that monopoles play a relevant role of 
the nonperturbative phenomena of QCD like 
the color confinement through monopole condensation.
In the abelian gauge,
the monopole appears from the hedgehog-like 
gauge configuration \cite{thooft,sug0}.
On the other hand, 
the nonabelian gauge theory in the Euclidean space 
has a classical solution called instanton.
Instantons are also believed to play an important role
in the QCD ground state.

In the abelian-dominant system, 
instantons seem to loose the topological basis for its existence,
since the instanton is a topological object in the nonabelian manifold.
However, even in the abelian gauge,
nonabelian components are not completely suppressed
around the topological defect, {\it i.e.} monopoles.
Therefore instantons are expected to appear only 
around the monopole trajectories in the abelian-dominant system.
In fact, the fourth component of 
the exact instanton solution in the singular gauge
becomes the hedgehog configuration as
\begin{equation}
A_4=\frac{i}{\phi}\sum_k 
\frac{{a_k}^2\ \mathbf{\sigma}\cdot ({\bf x}-{\bf x}_k)}{|x-x_k|^4}\ \ \ ,
\ \ \ \phi\equiv 1+\sum_k \frac{{a_k}^2}{|x-x_k|^2},
\label{eq:A4}
\end{equation}
where $x_k$ is the $k$-th instanton center position and $a_k$ is its size.
The Polyakov-like gauge \cite{sug1,sug2,araki,sug3},
in which $A_4$ is diagonalized,
provides a monopole on ${\bf x}={\bf x}_k$.
 Thus, the instanton center is inevitably penetrated by the monopole 
trajectory along the temporal direction in the Polyakov-like gauge.

\section{Instanton and anti-instanton system at finite temperature}

The exact solution of the Yang-Mills theory 
regards either instantons or anti-instantons.
However, since it seems natural that QCD has both of them,
we choose the R-ansatz \cite{shu},
\begin{equation}
A^\mu=-i\sigma^a\cdot
\frac{\eta_{-}^{a\mu\nu}\partial_\nu\phi_I
+\eta_{+}^{a\mu\nu}\partial_\nu\phi_A}
{\phi_I+\phi_A-1}
\label{eq:r-ansatz}
\end{equation}
with $\eta_{\pm}^{a\mu\nu}\equiv \varepsilon^{a\mu\nu 4}
\pm\delta^{a\mu}\delta^{\nu 4}\mp\delta^{a\nu}\delta^{\mu 4}$.
Here, $\phi_I$ and $\phi_A$ give 
instanton and anti-instanton profiles, respectively.
These are the same form as $\phi$
of Eq. (\ref{eq:A4}).
The finite temperature system is described 
by imposing the periodic boundary condition
on the Euclidean temporal direction,
with the period $2\pi T$,
where $T$ is the temperature of the system.
The profile function $\phi$ is modified as
\begin{equation}
\phi\longrightarrow
1+\pi T\sum_k\frac{{a_k}^2}{r_k}\cdot
\frac{\sinh(2\pi Tr_k)}
{\cosh(2\pi Tr_k)-\cos(2\pi T\tau_k)},
\label{eq:phiT}
\end{equation}
with $r_k\equiv |{\bf x}-{\bf x}_k|$ and $\tau_k\equiv t-t_k$.
In the Polyakov-like gauge,
the monopole trajectories appear on each center
of the instanton and anti-instanton 
as the case of the previous section.

\section{Numerical Simulation}

Next, we demonstrate the behavior of the monopole trajectories
in the above system by numerical simulations.
In Figs. 1 and 2, we show examples of the monopole trajectories 
in this system 
at the temperature $T=0$ and $T=1({\rm fm}^{-1})$, respectively.
In both cases, the monopole trajectories are
generated by 100 instantons and 100 anti-instantons
which are put on the $zt$-plane.
The black dot and the small circle 
denote the instanton and anti-instanton center, respectively.
In Fig. 1, 
the monopole trajectories are found to be highly complicated 
penetrating the centers of the instantons and the anti-instantons.
As the temperature goes high (Fig. 2), 
the monopole trajectories tend to become straight lines 
along the temporal direction and change their topology.

\section{Summary}
We have studied the connection between monopoles and instantons.
In the Polyakov-like gauge, 
monopoles appear on each instanton or anti-instanton center.
Therefore, the monopole trajectories 
inevitably penetrate through instanton centers.
We have studied the behavior of the monopole trajectories in 
the instanton and anti-instanton system at finite temperature.
At the zero-temperature,
the monopole trajectories tend to become highly complicated and very long,
which can be regarded as a signal of monopole condensation.
When the temperature goes high, 
the monopole trajectories tend to become simple,
and change their topology,
which may correspond to the vanishing of monopole condensation.

\begin{figure}[h]
\begin{minipage}[t]{13cm}
\psbox[width=5.3cm]{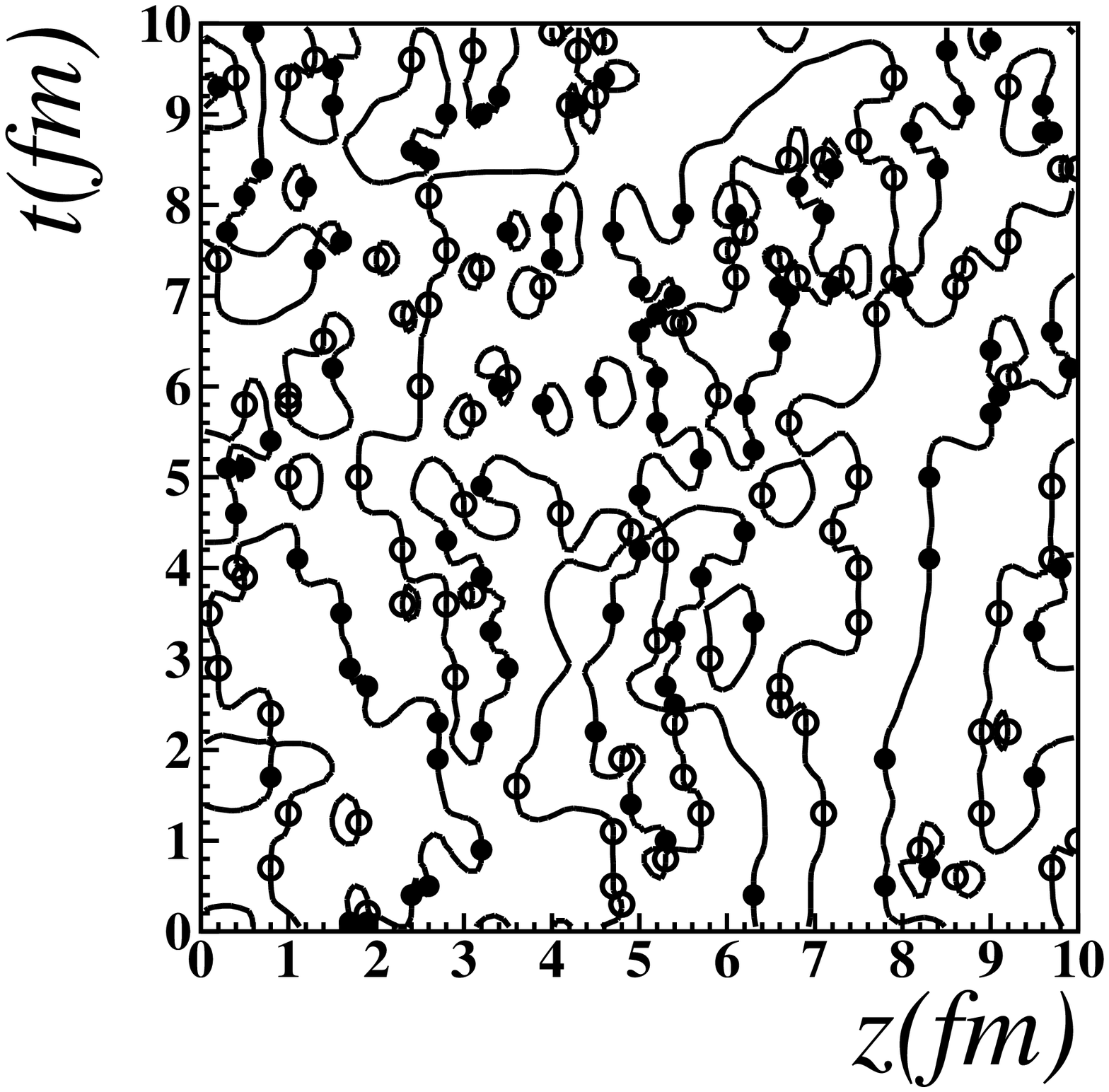}
\hspace{1cm}
\psbox[width=5.3cm]{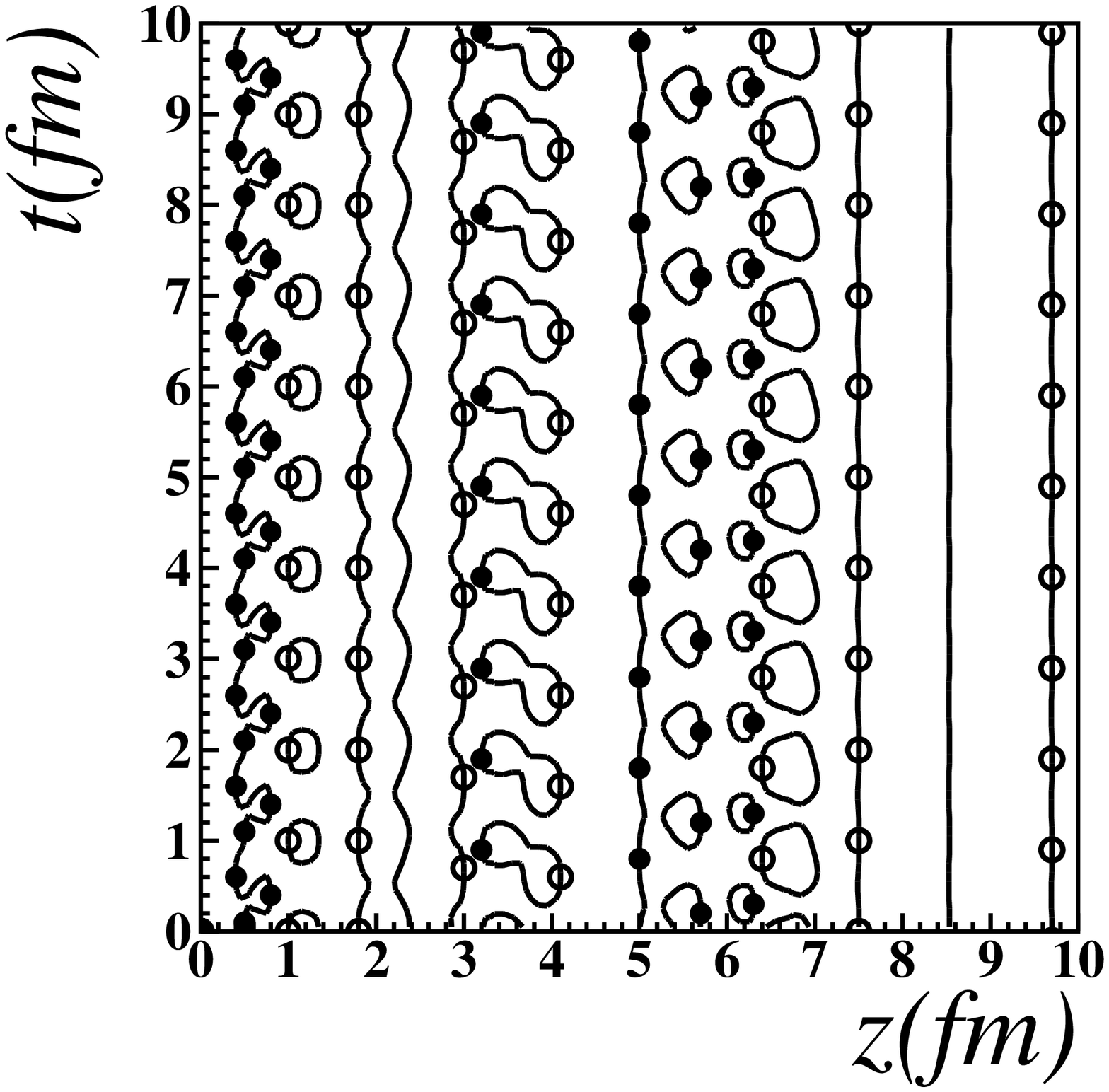}
\end{minipage}
\caption{The monopole trajectories 
in the instanton and anti-instanton system
at the temperature $T=0$.
The monopole trajectories are found to be highly complicated 
penetrating through the centers of the instantons and the anti-instantons.}
\caption{The monopole trajectories at $T=1({\rm fm}^{-1})$.
As the temperature goes high, 
the monopole trajectories tend to become straight lines 
along the temporal direction.}
\end{figure}

\end{document}